%Paper: astro-ph/9511102
%From: jasjeet@iucaa.iucaa.ernet.in (Jasjeet Bagla)
%Date: Wed, 22 Nov 95 15:28:18+050

\magnification = \magstep1
\baselineskip = 20  pt
\def\sss{\scriptscriptstyle}
\font\large cmbx10 at 12pt
\font\Large cmbx10 at 14.4pt
\font\small = cmr10 at 9 pt
\footline={\hfil}
\pageno=0
\newcount\equationno      \equationno=0
\newtoks\chapterno \xdef\chapterno{}
\def\eqn{\eqno\eqname}
\def\eqname#1{\global \advance \equationno by 1 \relax
\xdef#1{{\noexpand{\rm}(\chapterno\number\equationno)}}#1}

\centerline{\Large Crisis in Cosmology}

\centerline{\large Observational Constraints on $\Omega$ and $H_o$}

\vskip 2 true cm

\centerline{\bf J.S.Bagla$^1$, T.Padmanabhan$^2$ \& J.V.Narlikar$^3$}

\vskip 3.5 true cm

\centerline{\sl Inter-University Center for Astronomy and
Astrophysics,}

\centerline{\sl Post Bag 4, Ganeshkhind,}

\centerline{\sl Pune 411 007, INDIA}

\vskip 5 true cm

\centerline{ IUCAA 49/95,  November 1995}

\vfill

\hrule height1pt

\bigskip

\noindent\small $^1$  jasjeet@iucaa.ernet.in

\noindent\small $^2$  paddy@iucaa.ernet.in

\noindent\small $^3$  jvn@iucaa.ernet.in

\eject
\footline={\hss\tenrm\folio\hss}
\baselineskip = 17 pt

\centerline{\large Crisis in Cosmology : Observational Constraints
on $\Omega$ and $H_o$}

\bigskip

\hrule height 1pt

\bigskip

\bf

        Two decades ago, in an article in Nature, Gunn and Tinsley$^1$
had reviewed the then available data in cosmology to conclude: `` New
Data on the Hubble diagram, combined with constraints on the density
of the universe and the ages of galaxies, suggest that the most
plausible cosmological models have a positive cosmological constant,
are closed, too dense to make deuterium in the big bang, and will
expand for ever...". Thanks to new technology of observations and
fresh inputs from particle physics, cosmology has since advanced on
both observational and theoretical fronts.  The standard hot big bang
model has, if at all, become more deeply rooted in cosmology today
than in 1975.  It is therefore opportune that we take fresh stock
of the cosmological situation today and examine the observational
and theoretical constraints as they are now.  Not surprisingly,
some of the issues discussed by Gunn and Tinsley [ op. cit.]
continue to be relevant today whereas fresh ones have replaced
the rest. The purpose of this article is to carry out a similar
exercise in the modern cosmological framework. The bottom line in
this review is that despite the availability of the cosmological
constant as an extra parameter for flat Friedmann models, the allowed
parameter space for  such models has shrunk drastically. The observations
that we will consider here include the ages of globular clusters,
measurement of Hubble's constant, abundance of rich clusters of
galaxies,  fraction of mass contributed by baryons in rich clusters
and abundance of high redshift objects. We begin with a brief
description of the theoretical models in standard cosmology. For the
notation the reader may refer to standard textbooks$^{2,3}$.

\rm
\bigskip

\hrule height 1pt

\bigskip

\noindent{\bf The Standard Model } The standard scenario in
big bang cosmology assumes that at any given
time the universe is homogenous and isotropic when averaged over a
sufficiently large scale. The expansion of the universe is described
by the scale factor ``$a$'' that satisfies the following equation

$$\left({\dot{a} \over a}\right)^2 + {kc^2\over a^2} = {8 \pi\over 3}
 G \rho + {\Lambda \over 3}\eqn\qq $$
here $k=0, \pm 1$ represents curvature of the universe [  $k=+1$
represents a closed universe, $k=-1$ an open universe and
$k=0$ is the transition or the flat universe.], $\Lambda$ is
the cosmological constant and $\rho$ is the density of matter. We
can rewrite this equation for the present epoch as

$$H_0^2 + {kc^2 \over a_0^2} = \Omega_0 H_0^2 + \Omega_\Lambda H_0^2
\eqn\qq $$
where $H_0 = (\dot{a}/a )_{today}$ is the Hubble's constant, $\Omega_0$
is the density parameter for matter and $\Omega_\Lambda$ is the density
parameter contributed by the cosmological constant.

Radiation contributes very little to the energy density of the universe
at present, though it was the dominant contituent in the early universe.
For standard values of the parameters, matter decoupled from
radiation at a characteristic temperature of about $ 3000 K$. The
important relic from that epoch is believed to be the currently
observed microwave background with a precise blackbody distribution
with a temperature of about $2.7 K$.

It is also believed in the standard scenario that structure like
galaxies, clusters, etc. have formed out of growth of small scale
inhomogeneities via gravitational instability. Observations also
suggest that luminous ( baryonic) matter
forms only a small fraction of the total matter density, a much larger
contribution coming from ``dark'' matter, which is likely
to be nonbaryonic, noninteracting and collisionless.

The standard model described above has no clear mechanism for generating
small inhomogenities in the early universe. It is, however, possible to
come up with such a mechanism if one invokes the hypothesis that the
universe went through an inflationary phase at very high
redshifts. The models involving inflation generically lead to two
predictions: (i) The total density parameter $\Omega_0
+\Omega_{\Lambda}=1$ and (ii) The initial power power spectrum of
inhomogenities has the form $P_{in}(K) \propto K ^n$ with $n\simeq
1$. Over the years, the idea of inflation has undergone several
modifications to meet observational challenges, and it is now possible
to find a model in the market that will provide almost any value for
$\Omega_{total}$ and any form for $P_{in}(K)$.  For the sake of
definiteness we will only work with $n=1$ models. Observations of
microwave background radiation are consistent with the index $n$ being
equal to unity.
As the fluctuations grow the power spectrum gets modified at small
scales by different physical processes and this change is described by
a transfer function. We shall work with the transfer function
suggested by Efstathiou, Bond \& White$^4$, parametrised by $\Gamma\equiv
\Omega_0 h$. The power spectrum is normalised with the COBE DMR
observations$^5$ that give $Q_{rms-ps} = 20 \pm 3 \mu K$. Here
$Q_{rms-ps}$ is the amplitude of fluctuations in the quadrapole
inferred from fluctuations in the higher moments.

Here we study constraints on two models, namely those with (i) $\Omega_0 +
\Omega_\Lambda = 1 \; ; \;\; k=0$ and (ii) $\Omega_0 < 1 \; ; \;\;
\Omega_\Lambda = 0 \; ;
\;\; k=-1$. The first one is consistent with the inflationary models
though it requires an extreme fine tuning of the cosmological constant
which is contrary to the spirit of inflationary scenario. [ We shall
comment more on this later.] The second model may be thought of as an
``observer's model" in the sense that it tries to use what is known
observationally. The amplitude of fluctuations for open models is
obtained by rescaling the $\Omega_0 = 1$ model. Curvature effects are
not important as we are interested only in scales much smaller than
the curvature scale.

We next list the constraints arising from theory as well as
observations, giving a brief description of methods that are used to
obtain these and possible sources of errors for each constraint. Then
we merge constraints together to study the allowed regions in the
parameter space defined by the density parameter for matter ($\Omega_0$)
and Hubble's constant ($H_0$).

\medskip

\noindent{\bf Ages of Globular Clusters} : It is axiomatic that the
age of the universe must be larger than the ages of all its
constituents. Therefore ages of the oldest objects provide lower
bounds for the age of the universe. Stars in the globular clusters are
the oldest known objects in the universe. Ages of these stars are
computed by determining their mass, and by
observing metallicity and the position off the main sequence turnoff
point in the HR diagram. The uncertainities associated with these
determinations are now believed to be reasonably small and a fairly
accurate estimate for ages of stars can be obtained by this method.
Bolte and Hogan$^6$ compute the ages of stars in M 92 to be $15.8 \pm
2.1 {\rm Gyr}$.

The theoretical age of the universe can be readily computed given the
values of
$\Omega_0$, $\Omega_\Lambda$ and $H_0$. In figure 1 we have
plotted curves for $t_0 = 12, 15$ and $18 {\rm Gyr}$ (dashed lines);
the allowed region for any age lies below the corresponding
curve. Top frame shows these curves for flat models ($k=0$) and lower
frame shows the same curves for open models ($k=-1 \;\; ; \;\;
\Omega_\Lambda = 0$).

\medskip

\noindent{\bf Hubble's Constant} : We will use the parametrization
$h=H_0 / 100$ km s$^{-1}$ Mpc$^{-1}$. To measure $h$, we must measure
distance and recession velocity of a galaxy, or a group of
galaxies. Uncertainity in measurement of recession velocity of
galaxies comes mainly from their peculier motions. These can be
reduced by going to large recession velocities where the fractional
error arising from peculier velocities is small. Error in the distance
estimate depends upon the method that is used, and in general it
increases with distance. Distance indicators can be divided into three
classes, primary indicators like cepheid variables that can be
calibrated within our galaxy and therefore the uncertainity associated
with these is small. Secondary indicators like Tully-Fisher
relation are based on properties of galaxies as a whole and these have
to be calibrated with the help of galaxies to which distance has been
measured using primary indicators. This extra step involved tends to
increase errors in the computed distance. There is a class of
``physics'' based indicators which are independent of the entire
distance ladder, such as those using supernovae, the Sunyaev-Zeldovich
effect, etc.

Recent measurement$^7$ of distance to M 100 ( a galaxy in the Virgo
cluster) by the Hubble space telescope, with the use of the cepheid
period luminosity relation, gives the value $h=0.80 \pm 0.17$. This is
the ``local'' value of Hubble constant which may differ somewhat from
its global value. Turner, Cen \& Ostriker$^8$ and Nakamura \& Suto$^9$
have computed the probability distribution for the Hubble constant
given a local value. They show that values smaller than $h=0.5$ are
ruled out at $94\%$ confidence level. Global value of Hubble's constant
can be measured with methods like Sunyaev-Zeldovich effect. The
value determined by this method$^{10}$ for Abell 2218 is $h=0.65 \pm
0.25$. Sandage and Tamman, on the other hand consistently obtain
values of $h$ in the range $0.5 -0.6$ from a variety of methods [ See,
for example ref. 11].

In figure 1 we have plotted dotted lines bounding the region allowed
by the value
obtained for M 100 ( $0.63 < h < 0.97$) and also for $h=0.5$ as the lower
limit for the global value of the Hubble constant. If we assume that
$h=0.63$ [ The lowest value allowed by HST observations], then
$\Omega_0 = 1$ will require the age of globular clusters to be as
low as $10.6$Gyr. If $h=0.5$, we get  $t_0 = 13.3$Gyr. If the
age is greater
than $15$Gyr then we need $\Omega_\Lambda > 0.3$ for $h=0.5$ and
$\Omega_0 + \Omega_\Lambda = 1$. Thus a nonzero cosmological
constant is needed to allow for globular clusters as old as $15$Gyr.

\medskip

\noindent{\bf Abundance of Rich Clusters} : Mass per unit volume
contained in rich clusters can be estimated from the observed number
density of such clusters and their average mass. Clusters
are identified from x-ray observations by requiring the central
temperature to exceed $7 {\rm keV}$. Mass of these clusters can be
estimated by a variety of methods like assuming virial equilibrium and
using the velocity dispersion of galaxies, gravitational lensing
etc. One way of representing the observed number is to state
the contribution of mass in these clusters to the density parameter,
$\Omega_{clusters}^{obs}$. This number can be computed for any
theoretical model using the Press-Schechter method$^{12}$ and successful
models should satisfy the equality
$\Omega(>M_{clusters}) = \Omega_{clusters}^{obs}$, within the errors
of observations.

A comparison of observations with theory can also be carried out in
a more involved manner by converting the number density of clusters
into amplitude of density fluctuations at their mass
scale. This amplitude is then scaled to $8h^{-1}Mpc$ assuming power
law form for $\sigma$, the {\sl rms} fluctuations in density
perturbations. The index is chosen to match that expected in
the model being considered$^{13}$. The result is expressed as a
constraint on $\sigma_8$, the rms fluctuations at $8{\rm h^{-1} Mpc}$.

Errors in determination of $\Omega_{clusters}^{obs}$ are related
mostly to determination of mass for clusters. Various considerations
show that masses higher than those used for calculation are completely
ruled out, and in fact we may be overestimating the mass of these
objects. A change towards lower masses will tend to lower the allowed
values of $\sigma_8$. This will
shift the allowed region towards lower values of  $\Gamma=h\Omega_0$.

We have used observational constraints given by Viana \&
Liddle$^{14}$. For flat model the constraints are similar to those
given by White, Efstathiou \& Frenk$^{13}$. We have [ in figure 1]
plotted thick lines showing region
within one sigma of the mean. Thin lines show the bounds if
the uncertainity in COBE normalisation is taken into account. Top frame
shows these curves for the flat model while the corresponding curves for
open models are shown in the lower frame. These figures leave very little
room in the parameter space for open models. One may like to relax
(i.e., lower) the globular cluster ages and/or (lower) the Hubble
constant value somewhat to widen the allowed region, but all values
have to be pushed to their extreme limits for this purpose. The
allowed region for flat models is somewhat larger than in the $k=-1 \;
; \;\; \Omega_\Lambda =0$  case. Only three contraints have been used
so far and all of these are fairly robust.

\medskip

\noindent{\bf Baryon content of galaxy clusters} : Rich clusters of
galaxies like the Coma cluster have been studied in great detail. It
is possible to determine the
fraction of mass contributed by baryons to rich clusters by
assuming the Coma cluster to be a prototype. It is found that$^{15}$

$${M_{\sss B}\over M_{tot}} =
{\Omega_{\sss B}\over \Omega_0} \ge 0.009
+ 0.050 h^{-3/2}\eqn\qq$$
with $25\%$ uncertainity in the right hand side. This can be combined
with the value of $\Omega_{\sss B}$ determined from
primordial nucleosynthesis to further constrain $\Omega_0$.

Light nuclei form in the early universe as it cools from
a very dense high temperature phase. Relative abundance of different
elements is a function of $\Omega_{\sss
B} h^2$. The observed relative abundance of elements can be used to put
limits on this parameter [ See ref.16]. We use the values$^{17}$ $0.01
\le \Omega_{\sss B} h^2 \le
0.02$. [ There is no consensus
on the allowed range; therefore we are using a conservative set of
values.] By combining this value with the fraction of mass contributed
by baryons in clusters we can constrain $\Omega_0$. Plotted in figure
2 are the lowest and the highest bounds on matter density after
the uncertainity in the observations of fraction of mass contributed
by baryons has been taken into account. The permitted region lies to
the left of the curve. Reduction in uncertainity associated with these
observations can restrict the allowed region in parameter space very
effectively. Generalising to inhomogenous primordial nucleosynthesis
does not help as that tends to reduce the value of the baryon
density$^{18}$, leading to a tighter bound on $\Omega_0$.

Comparing figure 2 with figure 1, we notice that this constraint supplements
that given by the abundance of rich clusters for high values of $h$. For
small $h$ it is a stronger constraint and rules out more region from
the parameter space. A reduction in uncertainity in $\Omega_{\sss B} /
\Omega_0$ or a lowering of the upper bound on $\Omega_{\sss B} h^2$
can further reduce the allowed region. For example, if we use the mean
values of observations that combine to give this constraint, we will
rule out about one half of region that survives in the parameter space
in figure 1.

\medskip

\noindent{\bf Abundance of High Redshift Objects} : Existence of high
redshift objects like radio galaxies and damped lyman alpha systems
(DLAS) allows us to conclude that the amplitude of
density perturbations is of order unity at $M\simeq 10^{11} M_\odot$
at redshift $z=2$. We have plotted this lower bound in the top frame of
figure 2 for flat models and in the lower frame for open models. For flat
models, the curve runs almost
parallel to lines of constant age and thus provides {\it an upper
bound for the age of the universe}. If this constraint becomes stronger or
we discover globular clusters with age greater than $18 {\rm Gyr}$,
very little region will be left in the parameter space we are
considering. Similar results follow for open models.

A more rigorous calculation can be done along the same lines as that
described for abundance of clusters. However in the case of DLAS,
theoretically computed value of density parameter $\Omega(>M,z)$
should be greater than or equal to the observed value as not all
systems in that mass range host a DLAS. Observations of DLAS give us
the mean column density ($\langle\bar{N}\rangle$) of neutral hydrogen
and the number of DLAS per unit redshift ($dN/dz$). Using these and
the estimated neutral fraction for gas ($f_{\sss N} \sim 0.5$) we can
estimate the density parameter contributed by DLAS ( for more
details, see e.g. ref.19). It is also possible to compute the density
contributed by collapsed
objects at a given redshift using the Press-Schechter formalism. It is
important to ensure that collapsed objects of the relevant mass scale, in
a given model, are produced with the required abundance. It turns out
that this constrant is satisfied if DLAS are associated with masses
less than $10^{12} M_\odot$.

\medskip

\noindent{\bf Discussion} : These constraints rule out large regions and
the surviving region shrinks further or may even disappear if
observational uncertainity is reduced. In figure 3 we have
shaded regions that are allowed after taking all the constraints into
account. We have assumed that globular clusters are not older than
$12{\rm Gyr}$ and assumed $h>0.5$. A somewhat less conservative
interpretation of observations will lead to a much smaller allowed
region, shown here as cross hatched area.

This figure clearly shows that present observations rule out large
regions in the space of cosmological parameters. Flat models with
cosmological constant have a better chance of surviving as compared to
open models. We have not used other constraints coming from detailed
structure formation models like velocity fields, shape of the
correlation function etc. One reason for not considering them is the
large uncertainity associated with values derived from these. Another
is that we are able to rule out large regions in the parameter space
with only a handful of fairly robust constraints. Lastly, all the
constraints we use
can be scaled trivially if the observational uncertainities or values
of some input parameters change.

Allowed region in the parameter space can be widened if we allow tilted
spectra, i.e. spectra with the index of power spectrum $n\neq 1$. This
does not allow considerably greater freedom and we should keep in mind
that this puts strong limits on another parameter, namely the index of
the power spectrum. We have not specifically discussed any mixed dark
matter models as $\Omega_0 = 1$ models are ruled out by constraints
discussed above.

\medskip

\noindent{\bf Conclusions} This brief review highlights the new
developments in
cosmology since the review of Gunn and Tinsley$^1$.  Although the
constraints of ``age" have been with the big bang cosmology for
several decades, only now are they coming into focus, thanks to
the greater precision in the measurements of Hubble's constant
and an improved understanding of stellar evolution. Even allowing
for errors on both fronts,the conclusion today is inescapable
that the standard big bang models {\it without} the cosmological
constant are effectively ruled out.

       The constraints from structure formation, abundances of
clusters, primordial nucleosynthesis and high redshift objects
are all relatively recent ; but they additionally constrain the
models {\it even with the cosmological constant}. Indeed, with the
present understanding of extragalactic astronomy very little
parameter space is now left for the standard model with or without the
cosmological constant.

Finally, we would like to comment on the issue of ``fine-tuning". If
we take absence of fine-tuning to imply the dictum ``all dimensionless
parameters should be of order unity" then one would consider
$\Omega_{total}=1$ models as natural. [Any other model would require
fine-tuning of this parameter in the early universe, a difficulty
usually called ``flatness problem"]. By the same token one would have
insisted that $\Omega_{\Lambda} =0$. Such a model is clearly reuled
out by the observations. It is indeed hard to understand why the left over
cosmological constant is such as to exactly conform to the flatness
condition. As pointed out by Weinberg$^{20}$ this requires fine tuning
to one part in $10^{108}$. There have been attempts in the past to
invoke a dynamically evolving cosmological constant to circumvent this
difficulty ; however, none of these models have any compelling
features about them. At present, we must conclude that there is indeed
a crisis in cosmology.

\medskip

\noindent{\bf Acknowledgements } J.S.Bagla is being supported by Senior
Research Fellowship of CSIR India.

\bigskip

\noindent{\bf References}

\medskip

\item{1} Gunn, J.E. and Tinsley, B.M., Nature, 257, 454 ( 1975)

\item{2} Narlikar, J.V., Introduction to Cosmology, Cambridge ( 1992)

\item{3} Padmanabhan, T., Structure Formation in the Universe,
Cambridge ( 1993)

\item{4} Efstathiou, G., Bond, J.R. \& White S.D.M., MNRAS, 258, 1p (
1992)

\item{5} Gorski, K.M. et al., ApJ, 430, L89 ( 1994)

\item{6} Bolte, M. \& Hogan, C.J., Nature, 376, 399 ( 1995)

\item{7} Freedman, W.L. et al., Nature, 371, 757 ( 1994)

\item{8} Turner, E.L., Cen, R. \& Ostriker, J.P., Astron.J. 103, 1427
( 1991)

\item{9} Nakamura, T. \& Suto, Y., Preprint UTAP - 202/95 ( 1995)

\item{10} Birkinshaw, M. \& Hughes, J.P., ApJ, 420, 33 ( 1994)

\item{11} Saha, A. et al., ApJ, 438, 8 ( 1995)

\item{12} Press, W.H. \& Schechter, P., ApJ, 187, 452 ( 1974)

\item{13} White, S.D.M., Efstathiou, G. \& Frenk, C.S., MNRAS, 262,
1023 ( 1993)

\item{14} Viana, P.T.P. \& Liddle, A.R., Sussex preprint SUSSEX-AST
95/11-1 (1995)

\item{15} White, S.D.M., Navarro, J.F., Evrard, A.E. \& Frenk, C.S.,
 Nature, 366, 429 ( 1993)

\item{16} Olive, K.A. \& Scully, S.T., University of Minnesota preprint
UMN-TH-1341/95 ( 1995)

\item{17} Copi, C.J., Schramm, D.N. \& Turner, M.S., Science, 267, 192
( 1995)

\item{18} Leonard, R.E. \& Scherrer, R.J., Preprint OSU-TA-222/95 (
1995)

\item{19} Subramanian, K. \& Padmanabhan T., IUCAA-5/94 ( 1994)

\item{20} Weinberg, S., Revs. Mod. Phys., 66, 1 ( 1989)

\bigskip

\noindent{\bf Figure Captions}

\medskip

\noindent{\bf Figure 1} : This figure shows the constraints on the
density parameter
contributed by matter, $\Omega_0$, and the Hubble's constant $h$
arising from : (i) ages of globular clusters, (ii) measurements of
Hubble's constant, and (iii) abundance of rich clusters. Top frame
shows the constraints for a model with $k=0$, $\Omega_\Lambda \neq 0$
and $\Omega_0 + \Omega_\Lambda = 1$. The lower frame is for $k=-1$,
$\Omega_\Lambda = 0$ and $\Omega_0 \leq 1$ model. Lines of constant
age are shown as dashed lines for specified values of $\Omega_0$ and
$h$. Dotted lines mark the band enclosing value of local Hubble
constant ($0.63<h<0.97$) obtained from HST measurements. We have also
shown the assumed lower limit for its global value ($h=0.5$). Thick
unbroken lines enclose region which is permitted by the observed
abundance of clusters. Thin unbroken lines show the extent to which
this region can shift due to uncertainity in the COBE normalisation of
power spectrum. Note that these three constraints rule out large
regions in the parameter space. In particular, it is clear that the
$\Omega_0 = 1$ model is ruled out.

\noindent{\bf Figure 2} : This figure shows the constraints on the density
parameter contributed by matter, $\Omega_0$, and the Hubble's constant
$h$ arising from : (i) ages of globular clusters, (ii) measurements of
Hubble's constant, (iii) abundance of high redshift objects, and, (iv)
fraction of mass contributed by baryons in clusters and primordial
nucleosynthesis. Top frame shows the constraints for a model with
$k=0$, $\Omega_\Lambda \neq 0$ and $\Omega_0 + \Omega_\Lambda = 1$. The
lower frame is for $k=-1$, $\Omega_\Lambda = 0$ and $\Omega_0 \leq 1$
model. Lines of constant age are shown as dashed lines for specified
values of $\Omega_0$ and $h$. Dotted lines mark the band enclosing
value of local Hubble constant ($0.63<h<0.97$) obtained from HST
measurements. We have also shown the assumed lower limit for its
global value ($h=0.5$). Thick unbroken line is a lower bound on
permitted values of $h$ from abundance of high redshift objects. This
line depicts $\sigma(10^{11} M_\odot, z=2) =1$. Note that this
constraint implies that a $k=0$ universe can not be much older than
$18$Gyr. Dot-dashed lines mark the extreme upper limits allowed by
primordial nucleosynthesis and fraction of mass contributed by baryons
in clusters. For a given $\Omega_0$ allowed values of $h$ lie below
this curve ; conversly, for a given $h$, allowed values of $\Omega_0$
lie to the left of this curve. Uncertainities in all observations have
been included while plotting this curve.

\noindent{\bf Figure 3} : This figure summarises all the constraints
plotted in figures 1 and 2. Shaded region is permitted for $t_0 >
12$Gyr, $h>0.5$ and other constraints being satisfied. Cross hatched
area shows region with $t_0 > 15$Gyr and cluster abundance in the
allowed region without taking uncertainity in COBE normalisation into
account. If the uncertainities in the observations are pushed to the
extreme limits then the allowed parameter space corresponds to the
shaded region. A somewhat less conservative interpretation of
observations will lead to a much smaller allowed region, shown here as
cross hatched area.

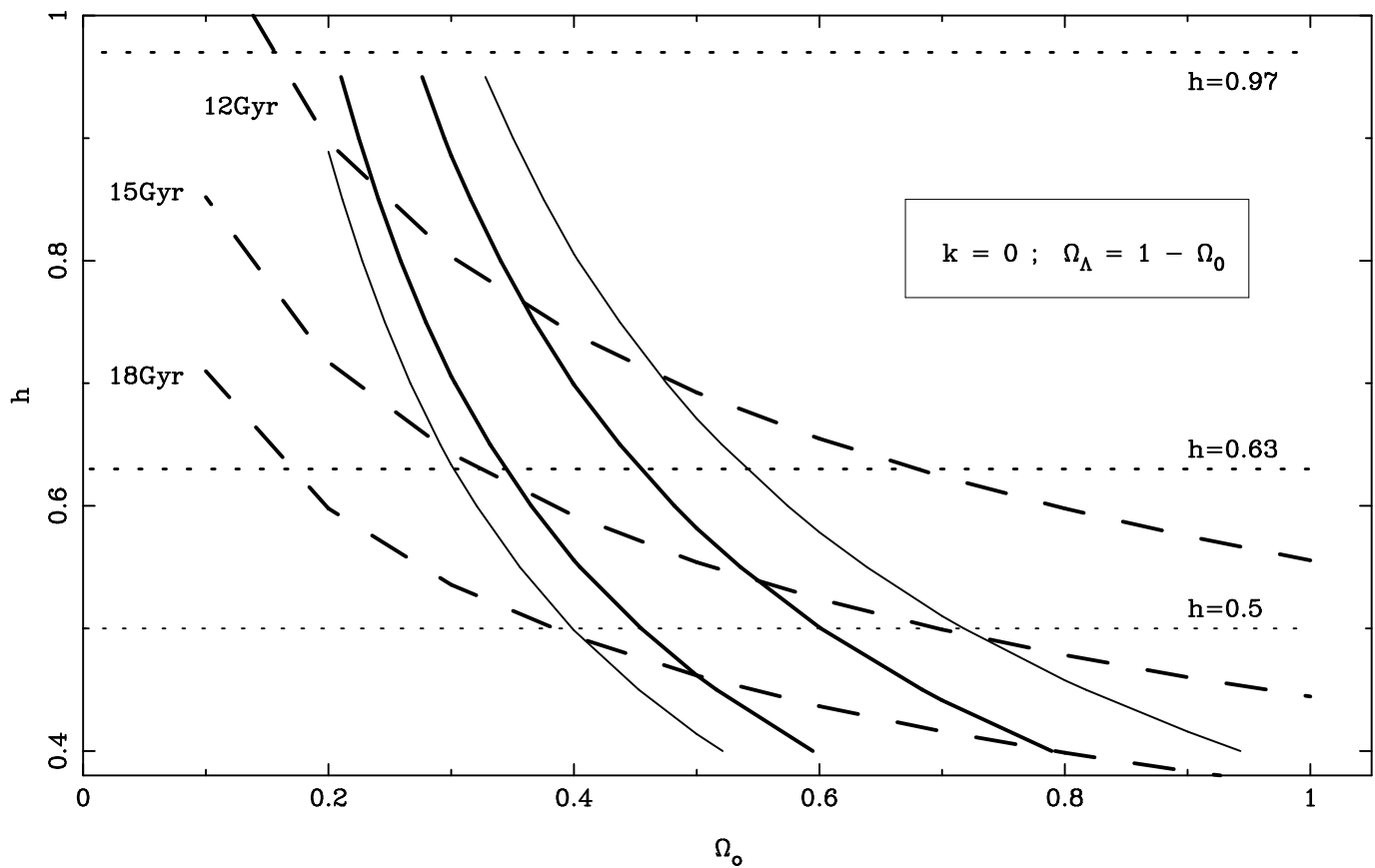
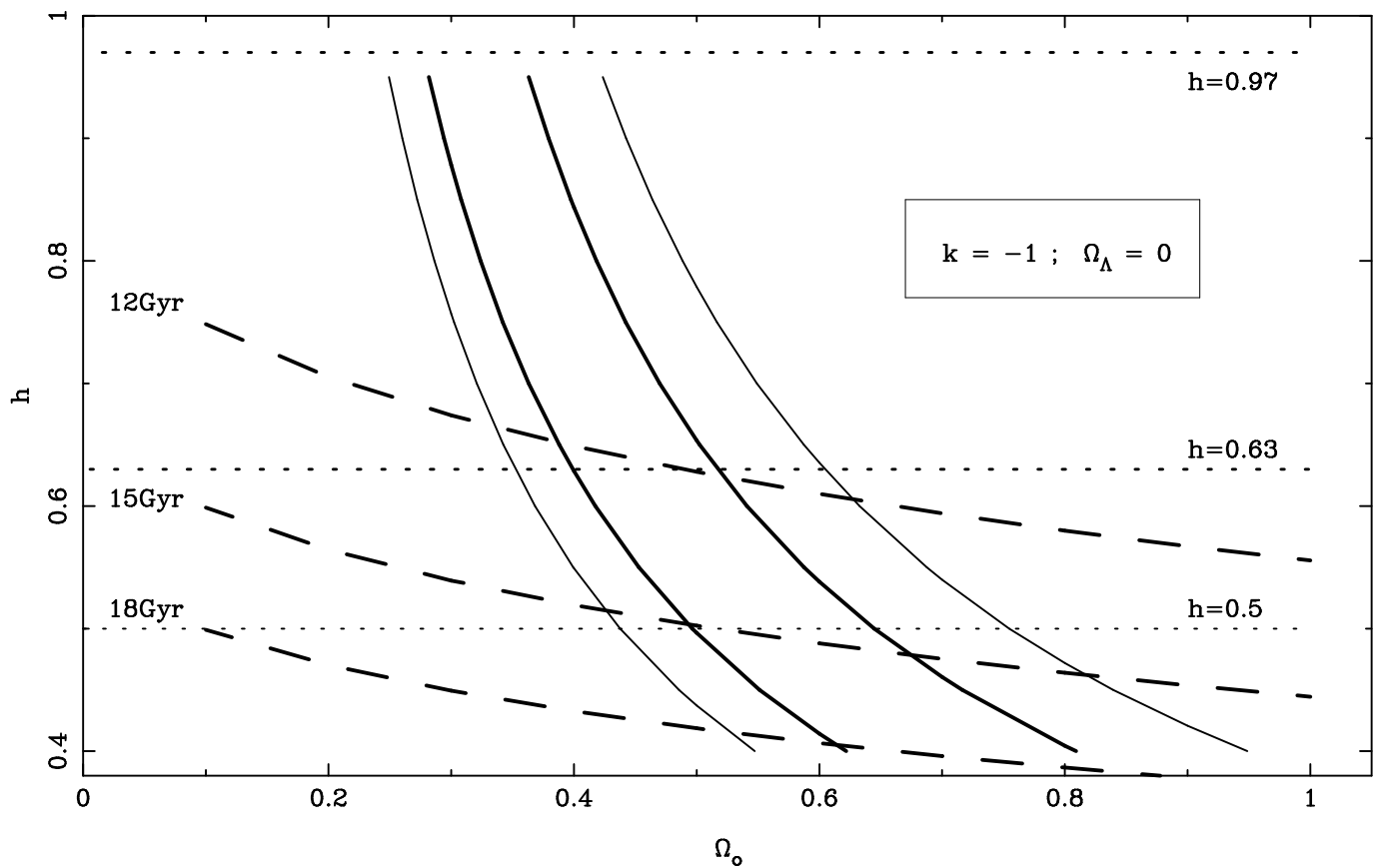
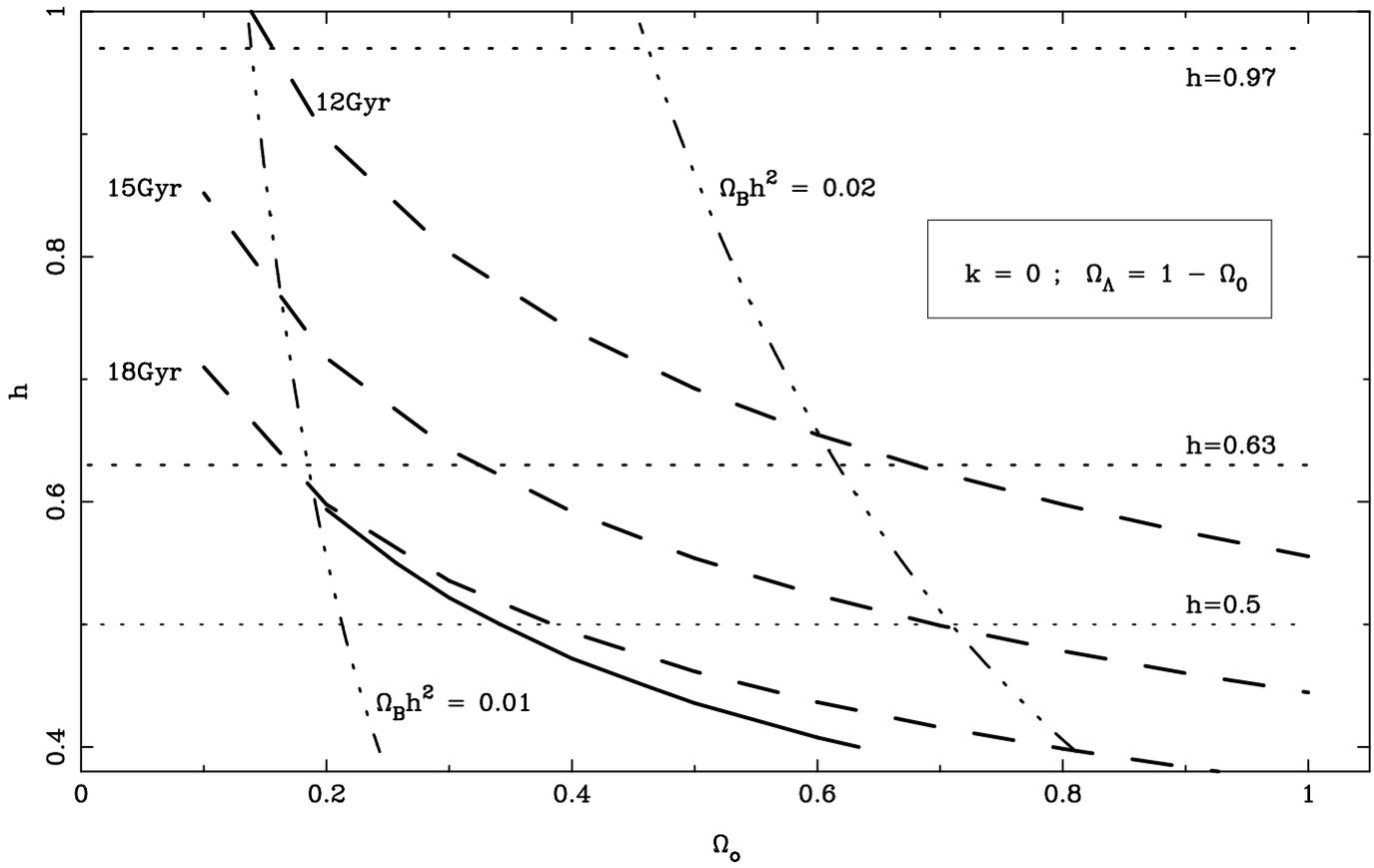
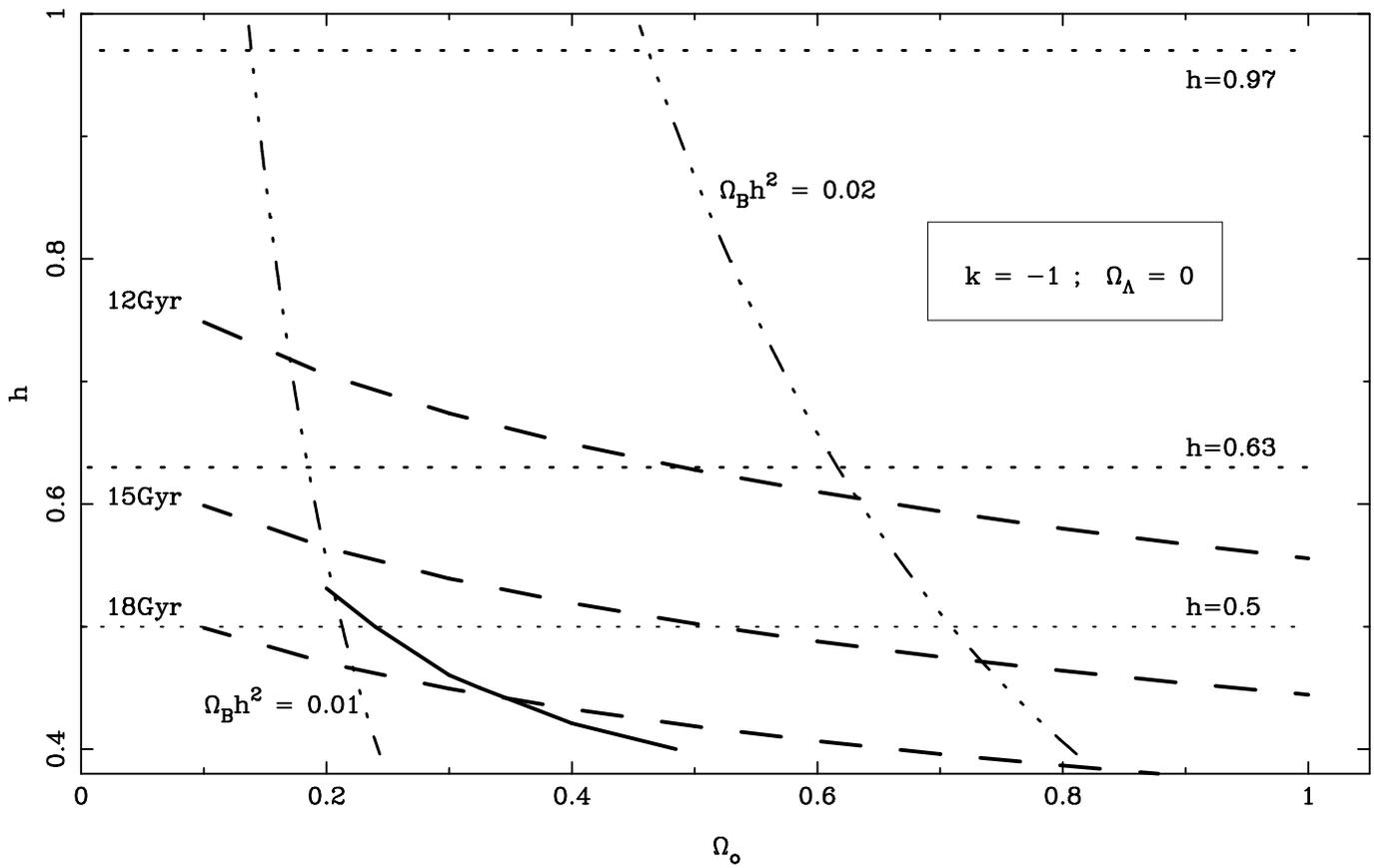
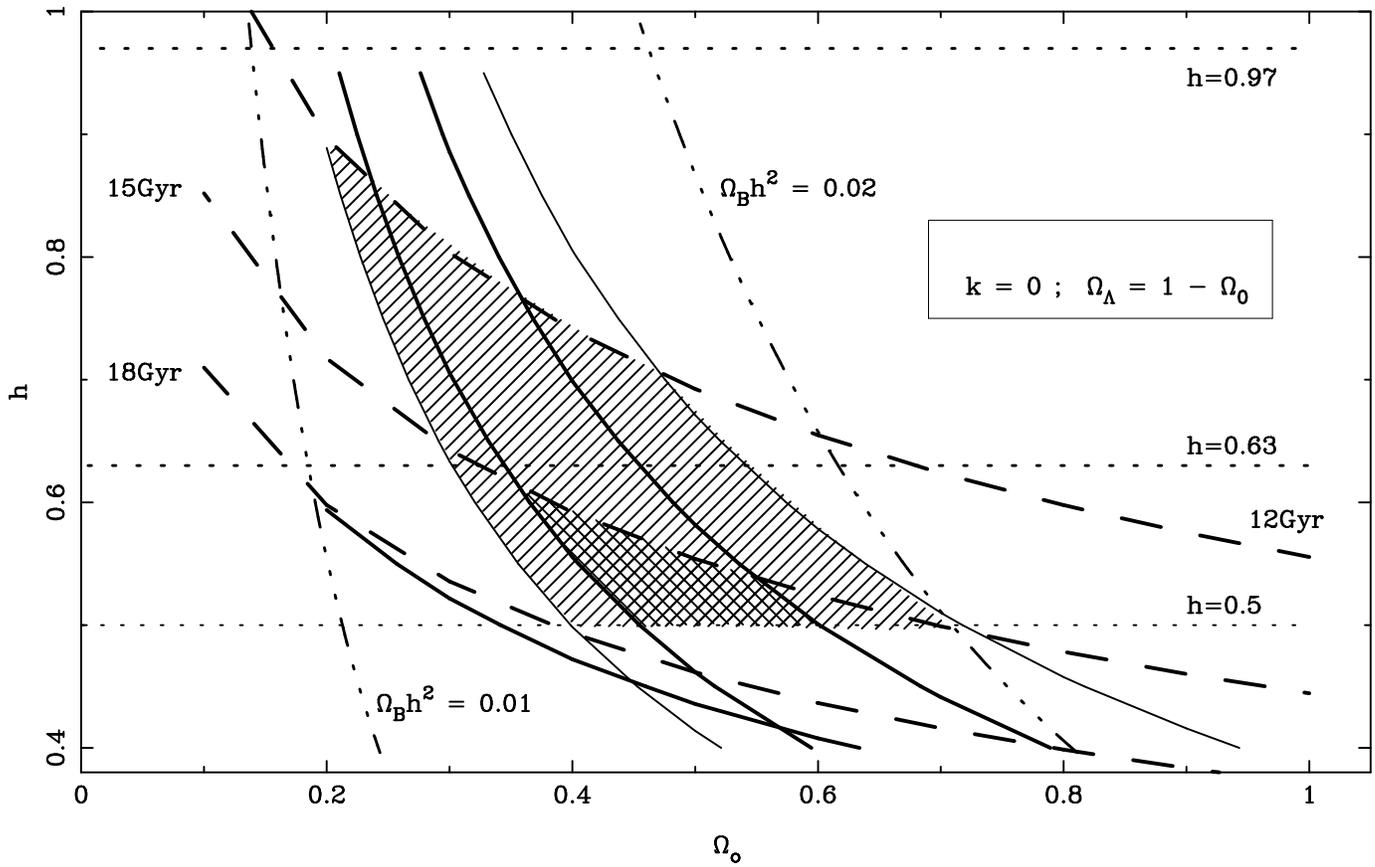
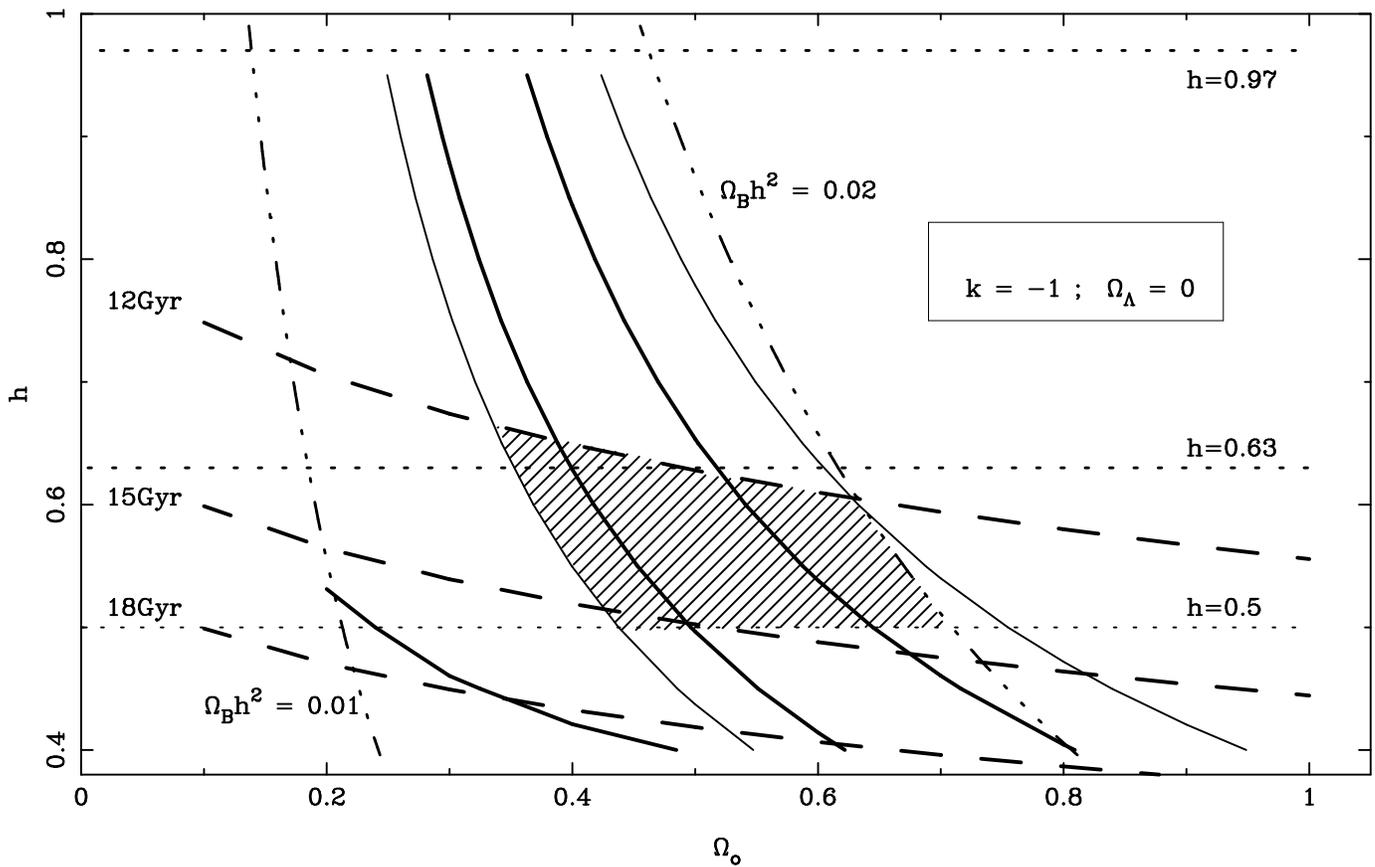
\bye